\documentclass[11pt]{article}

\usepackage[final]{acl}

\usepackage{times}
\usepackage{latexsym}
\usepackage{tikz}
\usepackage{url}
\usepackage{graphicx}
\usepackage{booktabs}
\usepackage{todonotes}
\usepackage{marginnote}
\usepackage{subcaption}
\usepackage{tcolorbox}    	
\usepackage{setspace}               
\usepackage{multicol}
\usepackage{xspace}
\usepackage{graphicx}
\usepackage{multirow}
\usepackage{xcolor}
\definecolor{headerbox}{HTML}{7BAFDE}
\definecolor{textbox}{HTML}{F5F5F5}
\definecolor{cblue}{HTML}{77AADD} 
\definecolor{cmint}{HTML}{44BB99} 
\definecolor{cgreen}{HTML}{AAAA00} 
\definecolor{corange}{HTML}{EE8866} 

\usepackage[T1]{fontenc}

\usepackage[utf8]{inputenc}

\usepackage{microtype}

\usepackage{inconsolata}

\usepackage{graphicx}

\newcommand{\rabpub}{\textsc{pub}\xspace}
\newcommand{\rabpriv}{\textsc{priv}\xspace}

\newcommand*\circled[1]{\tikz[baseline=(char.base)]{
            \node[shape=circle,draw,inner sep=2pt] (char) {#1};}}

\title{QUEST: A Query and Extraction System for Topics in Asylum Law Application Decisions}

\author{
  \textbf{Maria Vlachou\textsuperscript{1}},
  \textbf{Anna Murphy H\o genhaug\textsuperscript{2}},
  \textbf{Mohammad N. S. Jahromi\textsuperscript{2}},
  \textbf{Galadrielle Humblot-Renaux\textsuperscript{3,4}},
\\
  \textbf{Thomas Gammeltoft-Hansen\textsuperscript{2}},
  \textbf{Thomas B. Moeslund\textsuperscript{3,4}},
  \textbf{Desmond Elliott\textsuperscript{1,4}}
\\
\\
\\
  \textsuperscript{1}Department of Computer Science, University of Copenhagen\\
  \textsuperscript{2}Center of Excellence for Global Mobility Law, University of Copenhagen\\
  \textsuperscript{3}Visual Analysis and Perception Lab, Aalborg University,
  \textsuperscript{4}Pioneer Center for AI, Denmark
\\
  \small{
    \textbf{Correspondence:} \href{mailto:maria.vlachou@di.ku.dk}{maria.vlachou@di.ku.dk}
  }
}

\begin{document}
\maketitle
\begin{abstract}

Legal decisions on asylum applications consist of long, complex, and heterogeneous documents, covering narrative applicant interviews, original decisions, and additional supporting materials.
If an application is rejected, a critical question in processing an appeal is whether the credibility of the information in the original application was a factor that determined the original decision.
In this paper, we present the QUEST system (Query and Extraction System for Topics) to extract and identify factors relating to credibility assessments in two datasets of Danish asylum application appeals.
QUEST frames this problem as an information retrieval task, combining synthetic query generation, topic extraction, and relevance assessment to identify information related to credibility indicators in appeals board application materials.
In addition to standard retrieval evaluation metrics, we propose a new type of domain-specific assessments distinct from the traditional relevance to evaluate the performance of the tested systems with respect to credibility factors. 
In this way, we obtain insights about how well automatic methods can return answers for different types of indicators appearing in asylum appeals. 
Our results indicate that there is an increased challenge when estimating performance using credibility-based relevance assessments, thus pointing to the difficulty of the task. 

\textcolor{red}{\textbf{Warning:}} this paper contains derived text related to violence, persecution and trauma.

\end{abstract}

\section{Introduction}
One consequence of recent geopolitical  developments is an increase in asylum applications. In the European Union, as of November 2025, first time applications have risen by 62\% compared to the previous year\footnote{\url{https://ec.europa.eu/eurostat/web/products-eurostat-news/w/ddn-20260217-1}}, indicating an increased demand on the application and appeals processes. A major goal of our research is to understand the suitability of NLP techniques for the tasks of assisting asylum appeal applications at the Danish Refugee Appeals Board.
Specifically, language processing tools that can assist in automating the identification of critical information are expected to be valuable for dealing with this increased workload.
In this paper, we work with a dataset of appeals submitted to the Danish Refugee Appeals Board (RAB) from 2001 to 2025, which consists of heterogeneous materials, including a narrative transcript of an interview with the applicant, the original decision and justification, and additional materials, e.g. external evidence, photographs, and social media data.

The practical work of the Refugee Appeals Board is complex and legally involves making a forward-looking risk assessment based on limited evidence and dynamically evolving material on conditions in the applicant's home country.
A major factor in the final decision of the Appeals Board is whether the applicant is deemed {\em credible}.
Indeed, examining the \textit{credibility} of the applicant's testimony, i.e., the extent to which their narrative can be deemed as trustworthy, is common practice in asylum decision-making~\cite{dowd2018,herl2024psych,Jarlner03072026,liodden2020}, where domain experts rely on manual annotation \cite{hertz2025trans,hogenhaug2023nordic,rask2022data}.
Information related to credibility can appear at any point in the narrative interview transcript of either the original application or Appeals Board assessment.

Therefore, there is no straightforward way to identify when credibility assessments are a decisive factor without carefully reading the submitted materials. 
There are additional challenges that compound the difficulty of working with this type of data: (1) The material is {\em highly sensitive} and private, which constrains the use of models to those than can be run on highly-secure infrastructure.\footnote{The Ethical Statement at the end of this paper contains additional information on our use of this data.} 
(2) There are no external datasets that are likely to be useful for training, due to differences in legal frameworks between countries, and similar privacy concerns. 
(3) Some additional appeal materials, i.e, mobile phone messages, birth certificates, etc., are not readily processable by language models.
(4) The materials in the original application and appeals board materials are long, semi-structured narratives of an interview with the asylum applicant. 
The focus of this paper is mainly on how to handle the long and unstructured narratives. 
The data that we work with is available in two forms: (i) publicly available narrative summaries together with the outcome decision\footnote{\url{https://fln.dk/praksis/}}, denoted \rabpub, and a larger, highly sensitive set of complete materials relating to reassessment evaluations, denoted \rabpriv.

In order to identify text related to credibility factors, we frame our approach as an Information Retrieval problem over shorter chunks of the original documents.
We use automated tools to split the long documents into chunks, and identify topics that connect the chunks across different documents. 
From this perspective, we create queries using off-the-shelf query generation tools, given the document chunks and a domain-specific {\em codebook of credibility factors}. 
We draw inspiration from few-shot synthetic query-label generation methods in retrieval tasks~\cite{bonifacio2022inpars,daipromptagator}. 
Our evaluation is conducted using an LLM-as-a-judge~\cite{dietz2024workbench,gu2024survey,li2025generation,zheng2023judging}, given the limited human labour available for annotation. 
The query relevance assessments ({\em qrels}) come from a previously evaluated prompt~\cite{dietz2024workbench} that was validated on data from human-provided relevance assessment, where it displayed good correlations. 
We also directly evaluate the ability to retrieve documents related to credibility assessments, which we dub {\em crels}. 
This allows us to use the same data to assess different types of indicators, based on model performance using each type of produced labels.

Overall, this paper aims to answer the following research question: compared to human experts, how well does our proposed framework identify credibility indicators that appear in an Appeals Board report? 
Our findings indicate: (i) There is a large difference in performance between typical qrels-based assessment compared to credibility assessment of the tested retrieval systems. (ii) The retrieval systems are not tuned to retrieve documents related to credibility assessments. This could be changed to improve performance. (iii) Dense retrieval methods based on neural network representations do not consistently perform best, but re-ranking using a neural re-ranker consistently improves performance. (iv) If we apply a cut-off to the LLM-Judge, simulating stricter criteria for assessment, then the retrieval system performance decays until it collapses entirely.

In summary, our contributions are the following:


\begin{itemize}
    \item We propose QUEST, a \textbf{Qu}ery and \textbf{E}xtraction \textbf{S}ystem for \textbf{T}opics as an assitive tool to study the problem of identifying and extracting credilibity indicators from asylum appeal narrative interviews.
    \item We introduce {\em crels}, a domain-specific type of relevance assessment complementary to standard query relevance assesments. 
    \item We explore different styles of expert annotators by varying the threshold based on which they consider an item as relevant. 
\end{itemize}


\section{Related Work}

\paragraph{Credibility Assessment in Danish Asylum Law}
There is rich literature on credibility assessment in asylum decisions across disciplines and empirical contexts. Recently, a literature review showed how credibility assessment has led to methodological rifts and conceptual "fuzziness"~\cite{Jarlner03072026}. Traditionally, approaches to examining credibility have been qualitative~\cite{dowd2018,herl2024psych,liodden2020}. Nonetheless, better data access in countries such as Denmark has led to more systematic approaches, involving manual annotation and analysis \cite{hertz2025trans,hogenhaug2023nordic,rask2022data}, factors that explain Danish adjudication with classification methods~\cite{10.1145/3594536.3595155}, and LLMs to identify concepts for credibility annotation~\cite{bay2026managing}.
Danish asylum cases have been labeled for credibility (presence, and positive or negative) using a variety of LLMs and credibility descriptions~\cite{humblot-renaux-etal-2026-llms}. 

\paragraph{Legal NLP} LLM-as-a-judge is also used for legal tasks, either to evaluate RAG responses~\cite{enguehard-etal-2025-lemaj} or to construct domain-specific evaluation datasets using prompt clustering~\cite{raju-etal-2024-constructing}. In addition, existing approaches in legal NLP use document chunking to determine the relevance of exact spans of characters in legal texts~\cite{pipitone2024legalbench} and therefore use character-based retrieval. Unlike them, our task is more similar to short passage retrieval.

\paragraph{Topic Models}
Topic modeling refers to unsupervised approaches that identify themes in document collections ~\cite{alokaili2020automatic}. In earlier approaches, candidate labels retrieved from a pool are reranked with respect to their semantic similarity to topic terms~\cite{aletras2017labeling,aletras2014labelling,alokaili2019re}. More recently, ~\citet{alokaili2020automatic} proposed a model that takes a sequence of terms as input and generates a sequence of terms as labels. Our approach is more similar to the former, but uses a neural pipeline and an LLM for topic labeling.

\begin{figure*}[t!]
\centering
\includegraphics[width=1\textwidth]{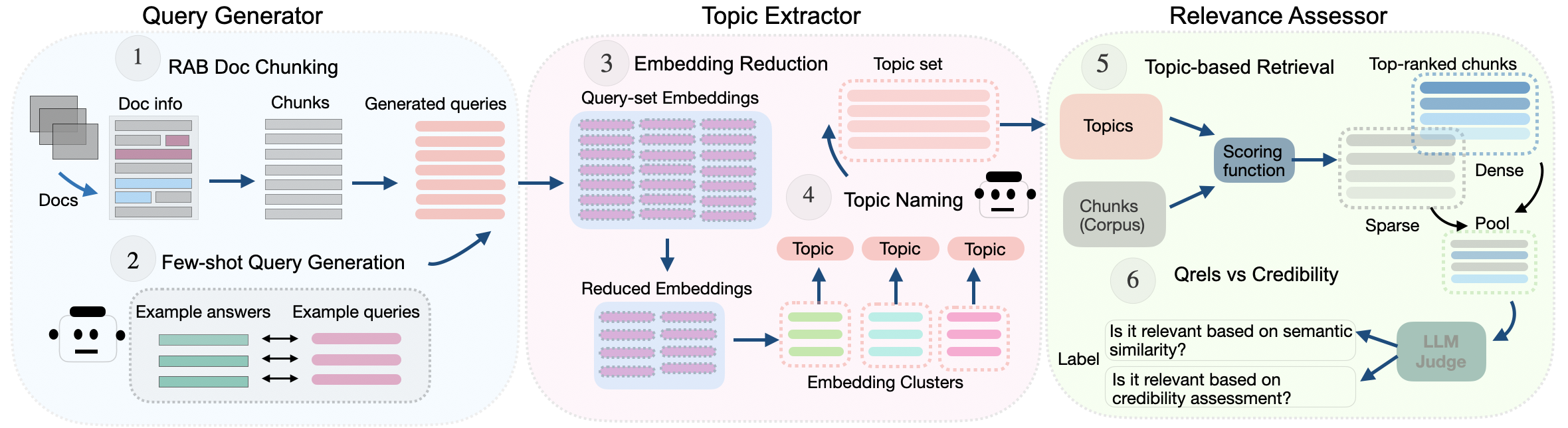}
\caption{Overview of the QUEST approach for extracting information from RAB files. In Step \protect\circled{1} each document of the RAB database is split into large sentences, and in \protect\circled{2} the Query Generator uses Few-shot examples to produce one query per chunk. In Step \protect\circled{3}, the resulting queries are embedded, clustered, and converted into topics by an LLM (Step \protect\circled{4}). Finally, the resulting topics are used to retrieve the top chunks (Step \protect\circled{5}), and consequently, in Step \protect\circled{6}, their relevance is assessed both in the standard and domain-specific way.}\label{fig:frame_fig}
\end{figure*}

\paragraph{Synthetic Relevance Judgments}

One approach to generate synthetic qrels using few-shot query generation with known relevant examples is InPars~\cite{bonifacio2022inpars}. It uses a language model to generate questions from documents, selects the top performing pairs as positive examples, and trains a reranker to estimate the relevance of documents to queries. ~\citet{daipromptagator} use few-shot query generation to create task-specific retrievers by fine-tuning dual encoders using in-batch negatives.~\citet{chaudhary-etal-2024-relative} go even further by generating irrelevant queries given relevant ones. 

\paragraph{LLM-based Relevance Assessment} There is increasing research using LLMs for relevance assessments~\cite{dewan2026true,faggioli2023perspectives,farzi2025criteria,rahmani2025judgeblender}. This extends to scoring text for various tasks using LLM judges~\cite{dietz2024workbench,li-etal-2025-generation}. Our LLM judge scores top-retrieved chunks by adjusting existing validated prompts~\cite{dietz2024workbench} to our domain-specific dataset.

\section{Methodology}

Here we present our overall framework. Our approach consists of a pipeline with three components, as shown in Figure~\ref{fig:frame_fig}. We first provide an explanation of how the task works and then we describe each component in more detail.

\paragraph{Problem Formulation}  Our task is to retrieve the most relevant document chunks related to a set of topics. More specifically, given a corpus $D$ of case files, a credibility codebook with legal definitions of credibility assessment, we create a set of chunks $C$ by splitting the documents into smaller segments. From this, we generate queries $Q$ for each chunk, and a set of topics $T$ formed by reducing the dimensions of the generated queries. Finally, we obtain relevance assessments of topic-chunk pairs, given the queries, to indicate the importance of the chunks to a number of indicators. 

\paragraph{Query Generator}
In Step \circled{1}, RAB documents are segmented into chunks composed of a couple of longer sentences, in line with examining the different parts of a RAB document. Then, Step \circled{2} generates queries for the chunks. In particular, similar to previous work on few-shot query generation~\cite{bonifacio2022inpars,carranza-etal-2024-synthetic,chaudhary-etal-2024-relative,daipromptagator} we use few-shot examples of external query-passage pairs to generate one query per chunk in the RAB data. 

One problem with few-shot example selection in our setting was that, unlike previous approaches that use traditional corpora with known relevant pairs, our dataset does not contain any labeled query-answer pairs. As a result, to enable few-shot generation, we decided to use question-answers from legal documents of Danish asylum law, and in particular a credibility codebook, a document used by legal experts in credibility assessment that provides definitions on credibility presence and indicators. Our example pairs are given in English,  the language in which the codebook is written. Our choice is based on the assumption that this document contains the known correct answers to a number of crucial legal questions of interest in Danish asylum law, and so they are treated as a set of ground truth answers; therefore, they can form positive examples. The few-shot generation prompt and the example pairs from the codebook are shown in Figures~\ref{fig:qgen_prompt} and~\ref{fig:few_shot_examples}, respectively.


\paragraph{Topic Extractor}
This component is responsible for producing a clustered version of the underlying semantic information contained in the generated queries. More specifically, 
Step \circled{3} takes the set of initial queries as input to extract topics. This is done using the BERTopic pipeline to obtain query embeddings, reduce their dimensions, and cluster the embeddings into topics. Then, Step \circled{4} uses Gemma3~\cite{gemmateam2025gemma3technicalreport} to give natural language names to each topic cluster. This allows the use of a crucial topic set that enables the method to automatically return the relevant content.  


\paragraph{Relevance Assessor}
This component is responsible for retrieving the relevant chunks and producing the different types of labels. The derived topic set is further used for the final steps of the process. As shown in Step \circled{5}, a set of sparse and dense retrieval methods are used to return the top-ranked chunks to the topics. 

Finally, Step \circled{6} involves relevance labeling to allow the calculation of evaluation metrics. To do this, we obtain a pool of the top-ranked chunks from all retrievers and use an LLM judge with Llamma3~\cite{grattafiori2024llama3herdmodels} to produce graded relevance labels. At this stage, we differentiate between two types of labels. First, the {\em qrels}, which traditionally provide relevance judgments of query-passage pairs. Second, we obtain labels more directly related to assessing the credibility of RAB files. In particular, instead of asking the LLM judge to consider whether the chunk adequately addresses the content of the topic, we ask whether the given topic is answered by assessing the credibility of the chunk content. These are domain-specific labels, which we call {\em crels}: relevance assessments with respect to credibility assessment. Figures~\ref{fig:qrels_prompt} and~\ref{fig:crels_prompt} show the prompts used to generate the qrels and crels, respectively.

In this way, we manage to provide labels for different tasks of interest using the same topic-answer pairs.  Our interest lies in the observed difference in rank metrics resulting from the different types of labels to determine the difficulty of each task.

\section{Experimental Setup}\label{sec:setup}

\paragraph{Retrieval Methods}
For first stage retrieval, we the following models: BM25~\cite{jones2000probabilistic} as a sparse retriever, two dense retrievers based on multilingual embeddings: E5~\cite{jiang2024e5vuniversalembeddingsmultimodal} and Qwen3~\cite{yang2025qwen3technicalreport}, and SPLADE~\cite{formal2021splade}, which generates sparse vector representations to bridge sparse and dense methods. Typically, dense retrieval models are more effective~\cite{khattab2020colbert,lin-etal-2021-batch,xiongapproximate}, as they are based on embedding models; still, they are computationally expensive. An alternative is to use a sparse model at first stage and use the power of a transformer to rerank the top-retrieved items. To test this, we use mono-T5~\cite{nogueira-etal-2020-document}, a standard practice to rerank BM25 results, while we also use reranking on Qwen3.



\paragraph{Evaluation Metrics}
Following standard TREC\footnote{\url{https://trec.nist.gov}} practices for measuring system effectiveness, we use MAP@100 and NDCG@10 for retrieval and reranking. In addition, we use MRR@10, which is the official metric of MS MARCO~\cite{bajaj2018msmarcohumangenerated}, a benchmark commonly used for evaluation. More details about the implementation of the full pipeline can be found in Appendix~\ref{sec:app_implementation}.

\section{Dataset}

Table~\ref{tab:data_stat} shows the statistics of our dataset sample and the resulting preprocessing of files into chunks. For \rabpriv, we use stratified random sampling by sampling 15 cases from each year. By keeping the chunks 2-3 sentences long, from an initial unequal document length (\rabpriv documents are much longer, as they contain the full narrative), we obtain a similar average chunk length between datasets.

\begin{table}[t!]
\small
\centering
\begin{tabular}{llllll}
\toprule
 & Samples   & Chunks & Words & Topics \\ \cmidrule(l){2-5} 
\rabpub     & 200    & 4,028  & 47.3  & 43     \\
\rabpriv    & 315    & 20,912 & 40.8  & 151    \\ \bottomrule
\end{tabular}
\caption{Statistics for the samples of the \rabpub and \rabpriv. The number of words is the average number of words per sample chunk.}\label{tab:data_stat}
\end{table}

\begin{figure}[t]
    \centering
     \begin{subfigure}[b]{1\linewidth}
    \includegraphics[trim={0cm 2cm 2cm 1cm},clip,width=1\columnwidth]{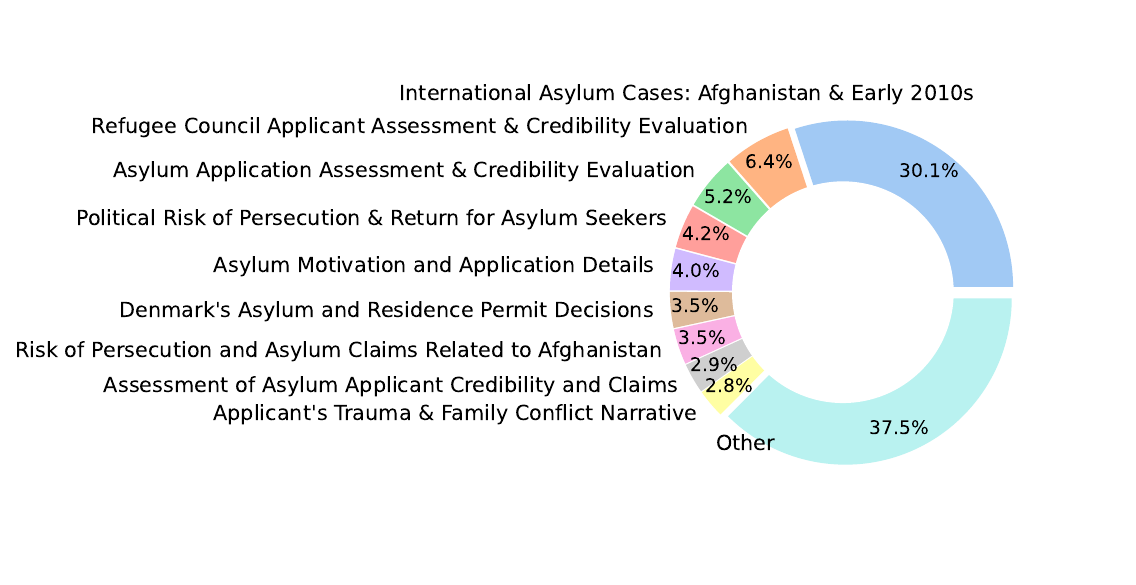}
     \caption{\rabpub}
     \end{subfigure}
     \begin{subfigure}[b]{1\linewidth}
     \includegraphics[trim={0cm 2cm 2cm 1cm},clip,width=1\columnwidth]{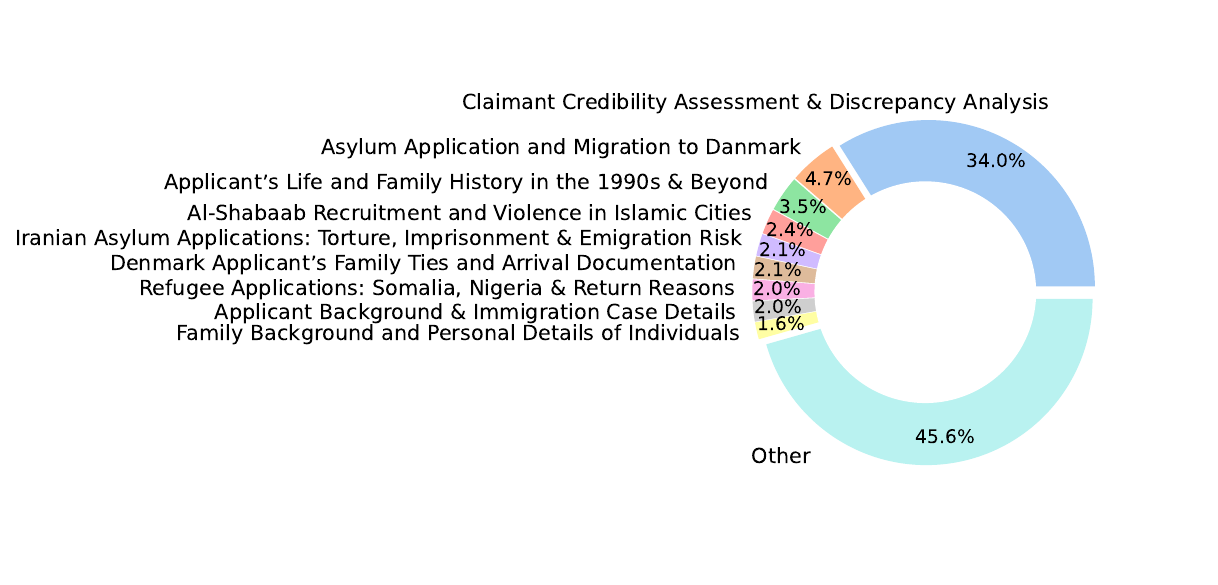}
     \caption{\rabpriv}
     \end{subfigure}
    \caption{Distribution of topics detected by the Topic Extractor for the \rabpub and \rabpriv datasets.}
    \label{fig:data_percent_all}
\end{figure}

Figure~\ref{fig:data_percent_all} shows the most frequent topics (semantic themes) contained in our data, as detected by the Topic Extractor. Indeed, the extracted topics refer to different types of information about the applicants. First, for \rabpub, the most prevalent trend refers to general information on cases from a particular country (Afghanistan) from last decade. This is country and year focused, but it could be relevant for both risk and credibility statements. The following few topics refer to either credibility or risk assessment, while the last one focuses on an applicant's trauma and family conflict. In this case, we observe that the most frequent topic more focused on the incidents within a specific time frame, relevant to a specific country of origin. 
In \rabpriv, there is a higher representation of credibility-based indicators, with the first topics reaching to 34\%. This finding is reasonable, since the private dataset contains longer case files, where credibility is more frequently mentioned across the document with respect to different aspects, and the discrepancies of a narrative are crucial for the outcome justification. Then, the following topics are more focused on specific countries and risk factors with connections to them, while the claimant's application details and ties to the country of entry are also examined. 

Table~\ref{tab:repr_queries} shows examples of representative generated queries from the chunks (used as input to BERTopic) that assigned to the corresponding topic (cluster). We observe that, although the information contained in the case files is highly sensitive, we manage to obtain questions from chunks that are quite generic in content, and can therefore easily be used to generate topics. Table~\ref{tab:repr_queries_3} contains more detailed examples and information about \rabpriv.



\begin{table*}[ht!]
\small
\centering
\begin{tabular}{p{7cm}p{8cm}}
\toprule
\multicolumn{1}{l}{Topic Name}                                                           & \multicolumn{1}{l}{Representative Queries}               
 \\[0.2ex] \midrule
International Asylum Cases: Afghanistan \& Early 2010s                & Refugee Name’s decision in asylum applications                                                     \\[0.5ex]
Applicant Assessment \& Credibility Evaluation        & Assess the credibility of a claim based on a relationship formed in Ukraine and fear of tribal violence \\[0.5ex]
Political Risk of Persecution for Asylum Seekers            & Risk of persecution by Iranian authorities due to political activity and family connections                     \\[0.5ex]
Asylum Motivation and Application Details                            & What is the applicant's asylum motivation?                \\[0.5ex]
Denmark’s Asylum and Residence Permit Decisions                & Will the asylum seeker be protected in Denmark under Section 7 of the Immigration law?                                                     \\[0.5ex]
Risk of Persecution and Asylum Claims Related to Afghanistan        & Assess the risk of persecution for an Afghan Shia Hazara refugee who left Afghanistan \\[0.5ex]
Assessment of Asylum Applicant Credibility and Claims            & Assess the credibility of an asylum applicant’s account                    \\[0.5ex]
Applicant’s Trauma \& Family Conflict Narrative                            & What happened to the applicant’s relationship with his father?                \\[0.5ex] 
Legal Reports and Documentation from Specific Years                            & What does the text say about the legal obligations of Immigration Service?                \\ 
\bottomrule
\end{tabular}
\caption{Representative Queries within the most frequent identified topics in \rabpub (relation between the Query Generator and the Topic Extractor). Some queries are slightly modified for privacy reasons.}\label{tab:repr_queries}
\end{table*}

\section{Results}
In what follows, we test system performance estimation by  measuring effectiveness using each type of relevance assessment and then examine the LLM responses compared to legal experts.

\begin{table}[ht!]
\small
\centering
\begin{tabular}{@{}lcccccc@{}}
\toprule
                    & &  \multicolumn{2}{c}{\rabpriv}                      & \multicolumn{2}{c}{\rabpub}       \\
\cmidrule(lr){3-4} \cmidrule(l){5-6}
                    & &  MAP        & NDCG                  & MAP        & NDCG         \\ 
\midrule
\multirow{4}{*}{\rotatebox[origin=c]{90}{qrels}} & BM25         & 0.166          & 0.475                   & 0.275          & 0.725                \\
 & SPLADE       & \textbf{0.430} & \textbf{0.676}   & 0.572          & 0.757              \\
 & E5         & 0.429          & 0.670                    & \textbf{0.589} & \textbf{0.768}      \\
 & Qwen3       & 0.397          & 0.515                   & 0.302          & 0.298              \\
\midrule
\multirow{4}{*}{\rotatebox[origin=c]{90}{crels}} & BM25       & 0.147          &  0.370                  & 0.226          & 0.435           \\
 & SPLADE       & \textbf{0.393} &  \textbf{0.500}  & 0.495          & 0.463          \\
 & E5          & 0.369          & 0.481                   & \textbf{0.520} & \textbf{0.465} \\
 & Qwen3        &  0.366         &  0.478                   & 0.300          & 0.236             \\
\midrule
\multirow{4}{*}{\rotatebox[origin=c]{90}{{label rels}}} & BM25   & 0.022          &  0.044                  & 0.055          & 0.117              \\ 
 & SPLADE & 0.037          &  0.079                   & 0.061          & 0.122           \\ 
 & E5    & 0.039          & \textbf{0.922}            & \textbf{0.072} & \textbf{0.146}  \\
 & Qwen3  & \textbf{0.059} & 0.114           & 0.066          & 0.099              \\ 
\bottomrule
\end{tabular}
\caption{First-stage retrieval performance according to query relevance (qrels), credibility relevance (crels), and label relevance (label rels). \textbf{Boldface} presents the highest performing system in each category.}\label{tab:retr_res}
\end{table}



\subsection{Estimating Performance with Different Types of Relevance Assessments}
Here, we examine retrieval and reranking focusing mainly in the differences between the qrels and crels settings, followed by variations in the extent to how strict an annotator can be.

\paragraph{Retrieval}


\begin{table}[ht!]
\small
\centering
\begin{tabular}{@{}lccccc@{}}
\toprule
                    & & \multicolumn{2}{c}{\rabpriv}                       & \multicolumn{2}{c}{\rabpub}       \\ 
\cmidrule(lr){3-4} \cmidrule(l){5-6}
         & &  MAP   & NDCG       & MAP   & NDCG   \\ \midrule
\multirow{3}{*}{\rotatebox[origin=c]{90}{BM25}} & qrels     & 0.137 & \textbf{0.538}    & 0.220  &  \textbf{0.776}  \\
 & crels     & 0.125      & 0.409             & 0.187 & \textbf{0.481}       \\
 & label\_rels &  0.020     & \textbf{0.063}              & 0.045  & \textbf{0.154}   \\ \midrule
\multirow{3}{*}{\rotatebox[origin=c]{90}{Qwen3}} & qrels     & 0.336 & \textbf{0.649}        & 0.139     & \textbf{0.688}   \\
 & crels     & 0.314      &   \textbf{0.497}          & 0.142 & \textbf{0.442}   \\
 & label\_rels & 0.055      &   \textbf{0.186}           & 0.065 & \textbf{0.211}   \\ \bottomrule
\end{tabular}
\caption{MonoT5 reranking performance of BM25 and Qwen3. \textbf{Boldface} represents improved performance compared to the corresponding first-stage performance.}\label{tab:rerank}
\end{table}

Table~\ref{tab:retr_res} shows the retrieval effectiveness of topic sets using each type of relevance assessment. Each group of rows corresponds to a different type of labels, and within each group, each row is a different retrieval method. As expected, we observe that BM25 is slightly worse than more advanced retrieval methods (more profoundly for MAP@100 and  less for NDCG@10). While dense retrieval methods are more effective in providing an accurate representation of the underlying semantic information, we don't observe any marked improvements from SPLADE to the multilingual embedding representation models. On the contrary, E5 and SPLADE provide comparable results, slightly higher than the rest.


In most cases, {\em crels} retrieval shows lower performance than the {\em qrels} setting, thus indicating the increased difficulty of the credibility assessment task. Indeed, qrels and crels have a different effect on performance estimation. This is made apparent by the observed Cohen's $\kappa$ values between qrels and crels, which are 0.04 for the public, and 0.02 for the private data, respectively. The agreement low between the relevance labels for the two tasks is expected, since the two tasks are genuinely distinct constructs the introduction of crels is justified.

While typically we do not compare performance between different types of qrels, the inspection of how performance changes when moving from standard to credibility assessment provides indications of how well we estimate performance in each case.



\paragraph{Reranking}
Table~\ref{tab:rerank} shows the reranking results from a monoT5 reranker. Each group of rows is a different retrieval method, and in each group, each row is a different type of relevance label. We use a standard BM25 >> monoT5 pipeline, while we also rerank an advanced method (Qwen3). We observe that MAP@100 does not improve after reranking the first stage results. Still, there is a slight improvement for NDCG@10 in most cases. Our main explanation for the lack of improvement for MAP is that we rerank the top-100 items, which is exactly how deep MAP goes, and therefore, part of the documents are not improved. However, NDCG considers items at a higher rank, which seem to be improved. Finally, the {\em crels} estimation is slightly lower than {\em qrels}, although the differences are smaller compared to Table~\ref{tab:retr_res}.

\paragraph{Can we benefit further from proxy available labels?}
On the way to reproducing state-of-the-art approaches for synthetic label generation~\cite{bonifacio2022inpars,daipromptagator} and inPars~\cite{bonifacio2022inpars}, the main obstacle is that we do not operate on the generated queries, but on the resulting topics. Also, unlike standard benchmarks, our dataset does not contain positive examples. Therefore, the closest available form of ground truth is the queries that contributed to each topic (BERTopic provides this link). Turning our attention to the last row in each group of Tables~\ref{tab:retr_res} and~\ref{tab:rerank}, we note that {\em label\_rels} do not improve performance estimation, although they are using this correspondence. Instead, in most cases, they result in lower accuracy compared to the LLM-judged settings. We believe this is partially due to the imbalanced topic counts, which do not allow a large range of qrels to less frequent topics.


\begin{figure}[t]
\centering
\begin{subfigure}[b]{1\linewidth}
     \includegraphics[clip,width=1\columnwidth]{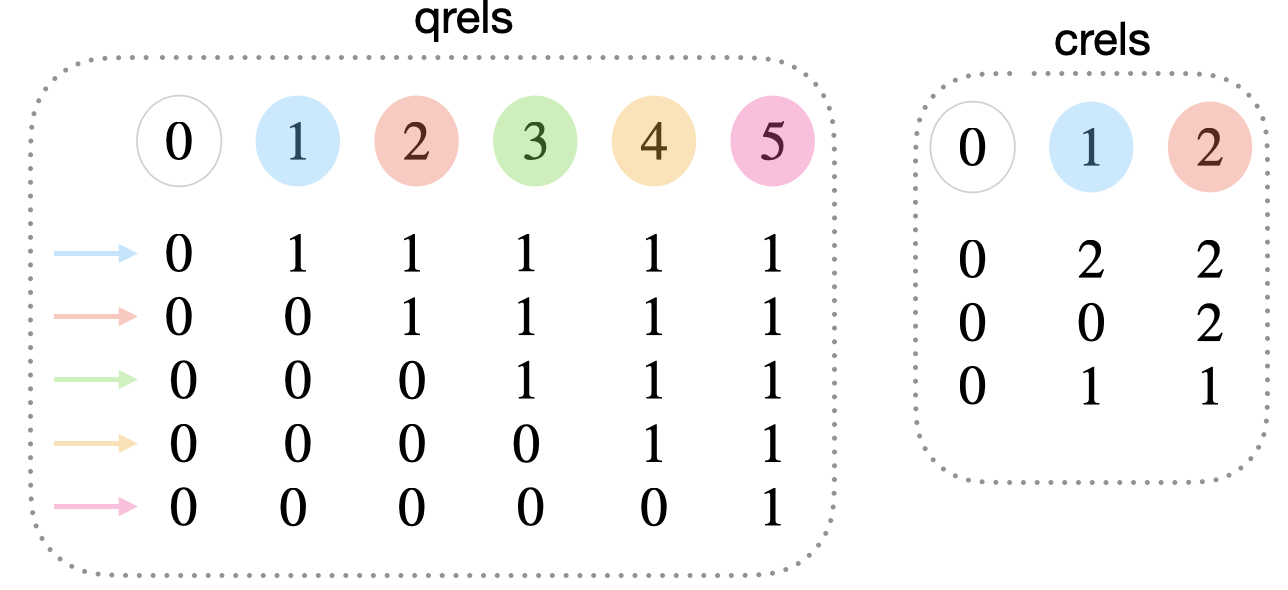}
     \end{subfigure}
\caption{Sketch of the binary evaluation threshold settings for relevance strictness (qrels and crels).}\label{fig:strict_sketch}
\end{figure}

\paragraph{Relevance Strictness}

\begin{figure}[t]
    \centering
     \begin{subfigure}[b]{0.49\linewidth}
     \includegraphics[width=\textwidth]{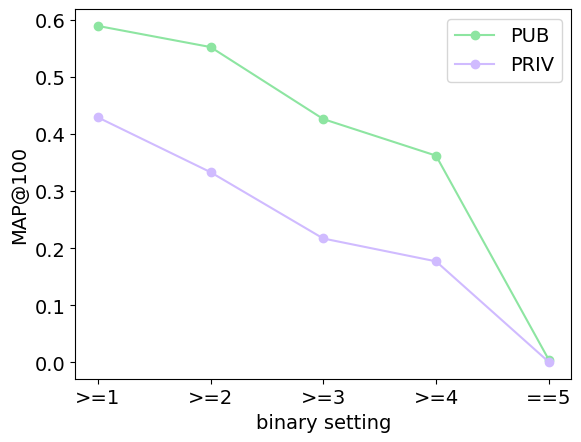}
     \caption{qrels}
     \end{subfigure}
     \begin{subfigure}[b]{0.49\linewidth}
     \includegraphics[width=1\columnwidth]{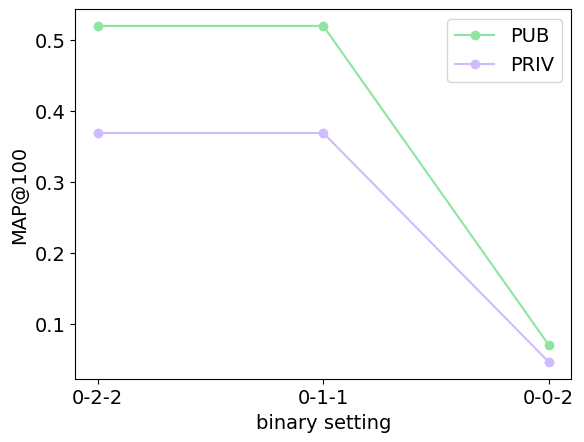}
     \caption{crels}
     \end{subfigure}
    \caption{\looseness -1 MAP@100 variations for the different qrels and crels strictness cutoffs for the RAB datasets.}
    \label{fig:qrels_var}
\end{figure}

Figure~\ref{fig:qrels_var} shows how retrieval performance varies with different relevance cutoffs. In particular, Figure~\ref{fig:strict_sketch} shows how we binarise the ratings to mimic more or less strict annotator styles in each case. For both datasets, we see a linear decrease in MAP@100 as we increase the point where a topic-chunk pair becomes relevant. We binarise the scoring of the LLM judge. Indeed, the more conservative cutoffs provide lower performance estimation. This implication should be taken into account in real annotation scenarios.


\subsection{Similarities and Differences between LLM Judge and Human Labels}

\begin{table}[t]
\small
\begin{tabular}{lcclll}
\toprule
 & Topics & Chunks   & $\kappa$ & $\alpha$ & $\rho$ \\ \cmidrule(l){2-6} 
\rabpub     & 5    & 5   & 0.130  & -0.095  & 0.543     \\
\rabpriv    & 5    & 5 & -0.010 & -0.333  & 0.333    \\ \bottomrule
\end{tabular}
\caption{Quantitative analysis of LLM judge sample.}\label{tab:llm_val}
\end{table}

\begin{table*}[ht!]
\small
\centering
\begin{tabular}{p{7cm}p{8cm}}
\toprule
\multicolumn{1}{l}{LLM Reason}                                                           & \multicolumn{1}{l}{Human Reason}               
 \\[0.5ex] \midrule
Score: 1 – The answer partially addresses the query, but has significant gaps or inaccuracies. It provides some context about the incident involving the claimant and the helper, but lacks details on the credibility assessment and discrepancy analysis, which are the main concerns expressed in the query.                & Score: 0 – The answer shows a description of motive. It does not concern a credibility assessment.                                                     \\\midrule
Score: 1 – The answer is mostly relevant and complete, but may have minor gaps or inaccuracies. It partially addresses the user's concern about discrepancy analysis, providing a plausible explanation for the applicant's inability to recall the threats. However, it does not fully clarify the discrepancies in the application.   & Score: 0 – Does not concern a credibility assessment (but excerpt from interview). However, the RAB finds it striking that the applicant does not remember details about previous threats (an indication of negative credibility finding). \\\midrule
Score: 1 –  The answer provides a clear assessment of the claimant's credibility and addresses potential discrepancies. However, it lacks some details about the specific grounds for the credibility assessment.                            & Score: 1 – The first sentence considers credibility (but vaguely). “Ansøgeren findes heller ikke at have sandsynliggjort” (The applicant has not made it probable …”)          \\
\bottomrule
\end{tabular}
\caption{Examples of scoring topic-chunk pairs using both the LLM judge and the human expert (the two columns provide judgments on the same pair). While in some cases the score might differ, the reasoning of the LLM and the human is very similar.}\label{tab:score_differences_main}
\end{table*}

Scaling the annotation with relevance labels to larger samples of credibility assessment is not feasible; therefore, LLM judges provide an option to annotate a wide number of samples. Here, we conduct qualitative and quantitative analyses to study the extent to which LLM labels converge with those provided by human experts. Indeed, we can estimate whether our LLM judge can mimic human assessments using a small subset. To achieve this, we use 5 topics from each dataset, and 5 chunks for each topic to compare the LLM with the annotations of an expert trained in credibility annotations on the topic-chunk pairs.

\paragraph{Quantitative Results} Table~\ref{tab:llm_val} shows the agreement levels using inter-annotator agreement with Cohen's $\kappa$ and Krippendorff's $\alpha$, and Spearman's $\rho$ for rank correlations. The observed values are in line with ~\citet{bavaresco-etal-2025-llms}, who observed low to moderate agreement in a wide variety of tasks. It should also be noted that our judge is asking a complex question in Danish, thus increasing the difficulty for an LLM to mimic human reasoning compared to a classic task of judging semantic relevance in English. The LLM often overestimates the credibility aspects of chunks by assigning partial instead of no presence, while the human annotator is consistently more strict, which makes it more similar to the strict setting (0-0-2) in Figure~\ref{fig:strict_sketch}(b). More details on the validation process can be found in Appendix~\ref{sec:app_validation}.

\paragraph{Qualitative Results}
The results in Table~\ref{tab:llm_val} motivate us to examine the reasoning humans and LLMs use to provide their domain-specific relevance scoring. Table~\ref{tab:score_differences_main} provides qualitative examples about the similarities and differences in assigned labels between the human and the LLM judge on the same pairs. It shows the reason they provided for assigning the label. In the first example, the human expert clearly finds the answer containing a motive and assigns 0, while the LLM judge also states that it has significant gaps or inaccuracies and does not appropriately address the main concerns of the query. Still, the LLM assigns partial relevance. The second example distinguishes between the experience of a human assessor and the LLM reasoning. In particular, while the LLM finds only minor inaccuracies and detects a discrepancy analysis, the human can recognize that while there is an indication (comment) about negative credibility from the board, this is not assessed in the interview. Finally, the third example shows agreement between the two raters, with the reasoning being similar.

Overall, we observe that the human rating is more strict with regards to whether credibility is assessed to answer a topic. This is because a human can best differentiate between the most and the least critical aspects of  complex text; in this case, it is more likely for the LLM to assign partial relevance. Still, there are cases where the LLM successfully mimics human performance.

\subsection{Implications of Findings and System Use}
We produced topics about indicators pointing to credibility as the consistency of information provided to an evaluator. A major finding is that using these topics, we can return relevant information in specific parts of a case file. This is an important point, as our goal is to identify which part of a case file contain key information to specific indicators. Indeed, top chunks can be from multiple case files. Then, returning to the case file structure, one can point to specific indicators using the chunk ids. Second, using domain-specific labels can inform us about the difficulty in returning content that uses credibility to answer a topic. To assess these topics, we use the information contained in available materials (legal assessments listing facts, evidence, and conclusions). Therefore, although further human bias may affect legal judgments, this is not present in the text and can therefore, not directly be assessed.  

Finally, the goal of QUEST is not to replace human experts in in credibility assessment, but to provide insights to experts on how to process information that could help them address the problem. Therefore, no deployment would take place without human oversight that would risk harmful decisions. Instead, we aim for a human-in-the-loop process that focuses on analysing and understanding existing decisions when time and resources are limited and is not directly involved in decision-making.

\section{Conclusions}

In our attempt to detect indicators in specific parts of  Danish Refugee Appeals Board appeals, we introduced a framework based on topic generation and topic-based retrieval to extract and identify information related to credibility assessments but also general information about the claimants.
Our analyses show that compared to human experts, our framework is overall less strict in identifying credibility indicators, while some topics result in answers more related to risk assessment. Therefore, our domain-specific relevance judgments are critical in determining whether credibility is assessed to provide a response to the task. 
Indeed, our topics present some data-derived indicators of RAB materials, and crels can distinguish performance estimation relevant to credibility compared to general labels. Explanations for annotations of human and LLM assessments should also be considered to assist experts in credibility assessment in identifying critical information.



\section*{Limitations}

Our proposed approach uses LLMs for various purposes: query generation, topic naming, and LLM-as-a-judge. While we use a different model to judge the relevance of top-ranked results than the one we used for query, we restrict ourselves to one LLM per task and only open weight LLMs. In addition, due to the nature of the data we have available, we base our conclusions on Danish data and the corresponding decision-makers' perspective. This inevitably intervenes in the process, and we would ideally have extended our results with datasets from other countries' asylum decisions if those were accessible to us.  Finally, we acknowledge the fact that the majority of our relevance assessments are not human derived, but at the same time, this is part of our contribution; to be able to move as close as possible towards a solution that approximates the process and task of a human's assessment (which would be impossible due to the amount of query-passage pairs) with the help of automated tools, and importantly, to have a judge that can differentiate between different types of labels.

\section*{Ethics Statement}
We used highly sensitive data from the Danish Refugee Appeals Board written in Danish, an under-represented language in
current language models and NLP datasets.
To work with the data, we have obtained ethical approval as part of research data registration. Finally, access to the dataset was made possible after signing a joint data controlling agreement that determines the purposes and means of processing of the dataset.
Given the challenges involved in biases~\citep{blodgett-etal-2020-language} and personal identifiable information anonymization~\citep{zent-etal-2025-piivot}, the system described in this paper is not currently deployed in any capacity.



\bibliography{custom}

\appendix

\section{Implementation Details}\label{sec:app_implementation}

In line with our privacy restrictions, for \rabpriv, all experiments were executed locally on an Apple M4 Pro with 24GB RAM. For \rabpub, part of the experiments (mainly the LLM-based relevance assessment scoring) were executed on an NVIDIA B200 with 192GB RAM. 
We implement the Query Generation, Topic Extraction, and LLM Judge steps  with \texttt{ollama}, which facilitates an easy LLM local implementation. Overall, regarding LLMs, we opted for open weight models for privacy reasons (specifically Gemma3~\cite{gemmateam2025gemma3technicalreport} and Llama3~\cite{grattafiori2024llama3herdmodels}) and for model sizes that can be run realistically on a local machine. 
Below, we describe our experimental choices with respect to each step of QUEST.  


\paragraph{Document Chunking}
We use the \texttt{spacy}\footnote{\url{https://spacy.io/}} sentence tokenizer with \texttt{langchain} Document\footnote{\url{https://reference.langchain.com/python/langchain-core/documents/base/Document}}. For \rabpriv, we use a sliding window of 8 sentences to enable an efficient way of obtaining a maximum content of information from the original document. This results in a total of 4,028 chunks for \rabpub and 20,912 for \rabpriv.

\paragraph{Few-Shot Query Generation}
We use a 3-shot example setting, where for each dataset, we use a fixed set of representative question-answer pairs from the credibility codebook. We choose the \texttt{gemma3:4b} model to generate one query per chunk, resulting in 4,028 and 20,912 queries, respectively.

\paragraph{Topic Extraction}
Using the generated queries from each dataset as input, we use the BERTopic\footnote{\url{https://maartengr.github.io/BERTopic/index.html}} pipeline with the following specifications: We use the \texttt{sentence-transformers/all-mpnet-base-v2}\footnote{\url{https://huggingface.co/sentence-transformers/all-mpnet-base-v2}} model to embed the original queries (which were generated in English), a UMAP model for nonlinear dimensionality reduction, and HDBSCAN for hierarchical clustering. We also use TF-IDF weighting to identify keywords per cluster. Finally, we use \texttt{gemma-3-4b-it-GGUF}\footnote{\url{https://huggingface.co/bartowski/google_gemma-3-4b-it-GGUF}} to provide names for each cluster, which corresponds to a given topic. This results in 43 topics for \rabpub and 151 topics for \rabpriv. Based on this, the extraction from RAB chunked files is now formed as cross-lingual retrieval.

\paragraph{Retrieval Methods}


We implement the retrieval methods with PyTerrier~\cite{macdonald2021pyterrier} by using our two datasets as custom collections. In particular, for BM25 and SPLADE, we generate a sparse index with \texttt{DanishSnowballStemmer}, while for E5 and Qwen3, we use the Huggingface checkpoints of Qwen3-0.6B~\footnote{\url{https://huggingface.co/Qwen/Qwen3-Embedding-0.6B}} and \texttt{multilingual-e5-small}~\footnote{\url{https://huggingface.co/intfloat/multilingual-e5-small}} and PyTerrier's FlexIndex. Finally, we use a multilingual monoT5 for reranking using the corresponding checkpoint~\footnote{\url{https://huggingface.co/castorini/monot5-base-msmarco}}. Our retriever choices are in line with recent suggestions about model performance, as reflected, for example, in this year's WSDM Cup\footnote{\url{https://wsdmcup-2026.github.io}} provided baselines for multilingual retrieval.

\paragraph{LLM Judge}
For both types of relevance assessments, we use the \texttt{llama3.1:8b} model. We opt for a model that is different from the one used for the query generation, and Llama has been found to be quite effective compared to other open LLMs for LLM-as-a-judge tasks \cite{bavaresco-etal-2025-llms}. To create the pool of each type of relevance assessment (general vs domain-specific), we merge the top-100 ranked results of each of the four retrieval models. To obtain the qrels, we adjust the prompt originally developed by~\citet{dietz2024workbench} for rubric-based evaluation of the top-ranked passages. The scale from 0 to 5 allows testing for different variations of relevance cutoffs. For the crels, we use a variant which allows for labeling credibility, non-credibility, and a partial assessment (scale from 0 to 2, 2 is maximum credibility).

\paragraph{Evaluation Metrics}
Following standard TREC\footnote{\url{https://trec.nist.gov}} practices for measuring system effectiveness, we use MAP@100 and NDCG@10 for retrieval and reranking. In addition, we use MRR@10, which is the official metric of MS MARCO~\cite{bajaj2018msmarcohumangenerated}, a benchmark commonly used for evaluation.

\section{Assessment Distributions}\label{sec:app_distributions}

\begin{figure}[ht!]
    \centering
     \begin{subfigure}[b]{0.80\linewidth}
     \includegraphics[width=\textwidth]{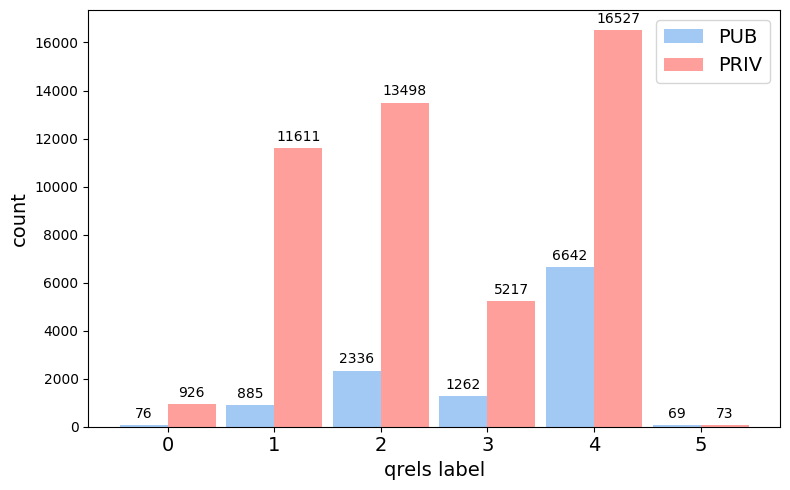}
     \caption{qrels}
     \end{subfigure}
     \begin{subfigure}[b]{0.80\linewidth}
     \includegraphics[width=1\columnwidth]{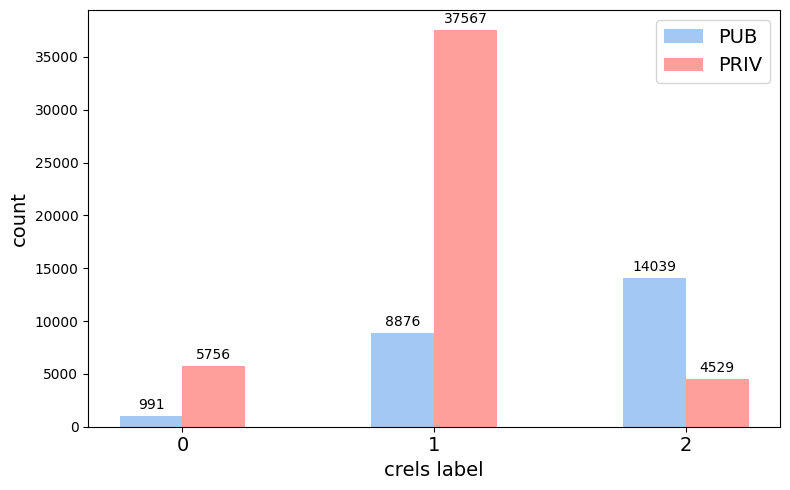}
     \caption{crels}
     \end{subfigure}
    \caption{Distribution of labels for the different types of relevance assessments.}
    \label{fig:rel_dist}
\end{figure}

This appendix shows the distribution of the different types of relevance assessments in the RAB data and in particular, in Figure~\ref{fig:rel_dist}, where we see the label distributions for both datasets. A first observation is that for qrels, the extreme label values receive very few counts for both datasets (this is even more notable for perfect relevance with label = 5). The most frequent labels are 2 and 4.

Moving on to the crels distribution, \rabpriv is much more unbalanced than \rabpub. The judge assigns cases to partially relevant with respect to credibility more frequently than the two absolute credibility labels. For \rabpub, a lot of cases are identified as using credibility to address topics (label = 2).

\section{LLM Judge Validation Details}\label{sec:app_validation}
In this appendix, we provide more details about the LLM Judge validation procedure. For the purpose of obtaining human labels, we engaged in discussions with domain experts, where we explained in detail how we approach the task. The main difference from how they normally analyse the text was that they consider the case file (or its summary) as a whole. We specifically asked about what aspects of text they consider to assign a credibility label. We organised dedicated sessions to explain the prompts we are using, the way we derive the topics, and how we judge relevance. 

During the annotation sessions, the assigned human expert was given the prompt with the instructions (same as the crels prompt given to the LLM judge), and the topic-chunk pairs were provided. After the end of the annotation task, we had an extended discussion about the most important aspects they considered for this particular task (when asked to judge the domain-specific notion of relevance). Their response was the following: They were looking for words they already know from the board assessment text, and they were searching for specific words that they intuitively know are related to credibility assessment. Note that we explicitly asked them if they were using a nugget-based approach, similar to~\citet{dietz2024workbench}, but their response was negative. Instead, they mentioned that the only purpose of spliting the text is to examine sentences separately to make the task easier. Therefore, we abstrain from using a nugget-based approach for our LLM judge.

Some further observations were the following: The domain expert noted that some topics do not refer to credibility. Indeed, these were the ones that mainly refer to risk assessment or to personal information about the claimant. Furthermore, we confirmed the clarity of the prompt and the instructions we provided, as they mentioned that the the description made their task less confusing. They were cases where they could also qualitatively differentiate between the two types or relevance we use, as they mentioned that for some pairs, they found relevance, but not in the domain-specific way. 



One final aspect of the human annotation was that the domain expert proposed that the LLM is sometimes not providing topic names in line with how a legal experts would define them. Therefore, we gave the opportunity to propose alternative topic names. This might result in slightly changing the overall meaning to increase its suitability to the domain. In Table~\ref{tab:new_topic_names}, we provide examples of such suggestions, which are useful for further studies. We note that some of the proposed names are indeed quite generic and could facilitate the identification of important information that refers to specific indicators.

\begin{table*}[ht!]
\small
\centering
\begin{tabular}{p{8cm}p{6cm}}
\toprule
\multicolumn{1}{l}{LLM Topic Name}                                                           & \multicolumn{1}{l}{Suggested Topic Name}               
 \\[0.5ex] \midrule
Applicant’s Life \&  in the 1990s \& Beyond                 & Asylum motive                                                     \\[0.5ex]
Asylum Application and Migration to Denmark        & Documents and questions\\[0.5ex]
Applicant’s Life and Family History in the 1990s             & Applicant’s family background and history                    \\[0.5ex]
Al-Shabaab Recruitment and Violence in Islamic Cities                             & Interview and country of origin information              \\[0.5ex] 
Refugee Council Applicant Assessment \& Credibility                            & Applicant motive and Credibility              \\[0.5ex]
Asylum Application Assessment \& Credibility Evaluation                            & Country Conditions and Credibility                \\[0.5ex]
Political Risk of Persecution and Return to Iran                            & Credibility and Risk                \\[0.5ex]
\bottomrule
\end{tabular}
\caption{Examples of topic reformulations based on the knowledge and experience of a human expert.}\label{tab:new_topic_names}
\end{table*}

\section{Prompts}\label{sec:app_prompts}
In this appendix, we include the set of prompts we use for our Query Generator and Relevance Assessor components of QUEST. First, we include the prompt given to the Query Generator using few-shot examples to generate one query per chunk using Gemma3.  This prompt, shown in Figure~\ref{fig:qgen_prompt}, is very similar to general-purpose benchmarks, where positive examples as used as few-shots to generate queries from passages.

\begin{figure}[htb]
\begin{tcolorbox}[
    colback=textbox,
    colframe=headerbox,
    colbacktitle=headerbox,
    fontupper=\fontsize{10}{12}\selectfont
]

\textcolor{cmint}{\textcolor{cmint}{\textbf{User prompt}}}\\
You are a helpful assistant trained in generating queries from text.
Given a text passage generate a query that exactly matches the semantic content of the passage.

Passage: \{\texttt{passage}\}\\

Here are some examples of relevant queries to passages.\\
Examples:\\
Passage: \{\texttt{passage}\} - Query: \{\texttt{query}\}\\

Only output the query without any additional text.
\end{tcolorbox}
\caption{%
Prompt used to generate queries for a given text passage.
}
\label{fig:qgen_prompt}
\end{figure}



In Figure~\ref{fig:few_shot_examples}, we see list the examples that we used for few-shot Query Generation from the credibility codebook. 
The easiest way to convert content from the credibility codebook into queries and passages was to identify critical aspects of how credibility assessment was defined in terms of presence, type, and the corresponding indicators. 
In this way, we managed to create a question that is answered by this content.

\begin{figure}[t]
\begin{tcolorbox}[
    colback=textbox,
    colframe=headerbox,
    colbacktitle=headerbox,
    fontupper=\fontsize{10}{12}\selectfont
]

\texttt{Passage 1}: The purpose of credibility assessment is to determine if the claimants narrative and account of events can be accepted as true (in whole or in part) or not. The focus is on the past and present facts asserted by the claimant. The credibility assessment relates to establishing facts, not whether these facts justify protection or whether there is a risk of persecution.\\
\texttt{Query 1}: provide a definition of credibility assessment\\\\
\texttt{Passage 2}: The overall credibility sentiment captures the decision-maker’s overall conclusion on whether the information provided by the claimant can be accepted as facts or not. It reflects the net outcome, not the presence of isolated inconsistencies or doubts.\\
\texttt{Query 2}: what is considered a sentiment of credibility\\\\
\texttt{Passage 3}: This refers to the assessment how plausible it is that the described events took place or people behaved as described, based on background information on conditions in the country of origin or the decision makers own assumptions of whether it is realistic that something took place or that someone in a given situation or under a given set of circumstances would behave in the described way. Plausibility can refer to specific events or behaviours, the chain of events, or the applicant's response thereto.\\
\texttt{Query 3}: what is meant by the perceived plausibility of past events and behaviours
\end{tcolorbox}
\caption{%
Examples of the few-shot passage--query pairs used in the query generation prompt.
}
\label{fig:few_shot_examples}
\end{figure}

In Figure~\ref{fig:qgen_prompt}, we details the prompt given to the LLM Judge of the Relevance Assessor zero-shot to generate relevance assessments on a graded scale using Llama3. The first prompt corresponds to the general purpose relevance assessments (qrels), adapted from ~\citet{dietz2024workbench}.

\begin{figure*}[htb]
\begin{tcolorbox}[
    colback=textbox,
    colframe=headerbox,
    colbacktitle=headerbox,
    fontupper=\fontsize{10}{12}\selectfont
]

\textcolor{cmint}{\textcolor{cmint}{\textbf{User prompt}}}

You are an expert evaluator in determining the relevance of text segments to queries. You will be given a user query and a system answer. Your task is to provide a 'total rating' scoring how well the system answer answers the user concerns expressed in the user query. In particular, you need to provide your score based on the following: Can the question be answered based on the available context? Give your answer on a scale of 0 to 5, where 1 means that the system answer is not helpful at all, and 5 means that the system answer completely and helpfully addresses the user query.\\

    Here is the scale you should use to build your answer:\\
    5: The system answer is highly relevant, complete, and accurate.\\
    4: The system answer is mostly relevant and complete but may have minor gaps or inaccuracies.\\
    3: The system answer is partially relevant and complete, with noticeable gaps or inaccuracies.\\
    2: The system answer  has limited relevance and completeness, with significant gaps or inaccuracies.\\
    1: The system answer is minimally relevant or complete, with substantial shortcomings.\\
    0: The system answer is not relevant or complete at all.\\

    Provide a complete and concise answer to the question based on the context as follows:\\

    Query: {row['user query']}\\
    Response: {row['system answer']}\\

   Return ONLY a JSON object with:\\
    - "score": (int between 0 and 5)\\
    - "reason": (brief justification)
    \end{tcolorbox}
\caption{%
Prompt used to perform standard relevance assessment.
}
\label{fig:qrels_prompt}
\end{figure*}

Finally, we include the prompt we use, again with Llama3, to produce the domain-specific relevance assessments (crels), as shown in Figure~\ref{fig:crels_prompt}.

\begin{figure*}[htb]
\begin{tcolorbox}[
    colback=textbox,
    colframe=headerbox,
    colbacktitle=headerbox,
    fontupper=\fontsize{10}{12}\selectfont
]

\textcolor{cmint}{\textcolor{cmint}{\textbf{User prompt}}}

You are an expert legal annotator specializing in Danish asylum law. \newline

Your role is to examine text segments of Flygtningenævnet (Refugee Appeals Board) decisions and identify whether credibility assessment determines their relevance to queries. Credibility assessment (troværdighedsvurdering) refers to evaluations and examinations of whether the asylum seeker's account is believable. You will be given a user query and a system answer. \newline 

Your task is to provide a 'total rating' scoring how well the system answer answers the user concerns expressed in the user query by also considering the credibility of the content with respect to the query. In particular, you need to provide your score based on the following: Is the given question answered by assessing the credibility of the available context? \newline

Give your answer on a scale of 0 to 2, where 0 means that the system answer is not examining credibility at all, and 2 means that the system answer is fully taking credibility into account to address the user query.\\
    Here is the scale you should use to build your answer:\\
    2: The system answer  uses credibility assessment to answer the query.\\
    1: The system answer partially uses credibility assessment to answer the query.\\
    0: The system answer does not use credibility assessment to answer the query at all.\\

    Provide a complete and concise answer to the question based on the context as follows:\\

    Query: \texttt{user query}\\
    Response: \texttt{system answer}\\

   Return ONLY a JSON object with:\\
    - "score": (int between 0 and 2)\\
    - "reason": (brief justification)

\end{tcolorbox}
\caption{%
Prompt used to assess identify whether credibility assessment determines their relevance to queries.
}
\label{fig:crels_prompt}
\end{figure*}

\section{Additional Results}\label{sec:app_additional_res}
In this appendix, we include complementary results regarding of all the components of our QUEST framework. Specifically, Table~\ref{tab:repr_queries_3} shows the topics and representative queries from the rest of \rabpriv. In some cases, we change some details, such as the name of a city or more specific identifiable details to more general terms, in order to stay in line with our privacy regulations. 

Finally, Tables~\ref{tab:retr_res_2} and~\ref{tab:rerank_2} show the retrieval and reranking results for both datasets using MRR@10 as the evaluation metric. Especially from the first-stage retrieval results, we observe that MRR@10 is not an appropriate measure for effectiveness in the {\em qrels} setting, as it mainly considers the relevance of the top-ranked item, and in our case it provides overestimated results.

\begin{table*}[ht!]
\small
\centering
\begin{tabular}{p{7cm}p{8cm}}
\toprule
\multicolumn{1}{l}{Topic Name}                                                           & \multicolumn{1}{l}{Representative Queries}               
 \\[0.5ex] \midrule
Credibility Assessment \& Discrepancy Analysis                & What are the conditions under which an applicant can contact the Immigration Office to search for relatives?                                                     \\[0.5ex]
Asylum Application and Migration to Denmark        & Discrepancies in claimant’s account regarding contact with police regarding spouse’s location\\[0.5ex]
Applicant’s Life and Family History in the 1990s             & Applicant’s family background and history                    \\[0.5ex]
Al-Shabaab Recruitment and Violence in Islamic Cities                             & Organisation’s contact with the claimant’s children               \\[0.5ex] 
Iranian Asylum Applications: Torture, Imprisonment \& Emigration Risk                            & Risk of imprisonment and torture by Iranian authorities                \\[0.5ex]
Denmark Applicant’s Family Ties and Documentation                           & Applicant arrived in Denmark on XX without valid travel documents and has no family ties to Denmark                \\[0.5ex]
Refugee Application: Somalia, Nigeria Return Reasons                           & Childhood in city X, Somalia                \\[0.5ex]
Applicant Background \& Immigration Case Details                           & Applicant’s family background and relocation history                \\[0.5ex]
Family Background and Personal Details of Individuals                            & X’s personal information: name, date of birth, and ethnicity                \\[0.5ex]
\bottomrule
\end{tabular}
\caption{Representative Queries within the most frequent identified topics in \rabpriv (relation between the Query Generator and the Topic Extractor). Some queries are slightly modified for privacy reasons.}\label{tab:repr_queries_3}
\end{table*}

\begin{table}[ht!]
\small
\centering
\begin{tabular}{@{}lccc@{}}
\toprule
                    & &  \multicolumn{1}{c}{\rabpriv}                      & \multicolumn{1}{c}{\rabpub}       \\
\cmidrule(lr){3-3} \cmidrule(l){4-4}
                         & &  MRR@10                & MRR@10   \\ 
\midrule
\multirow{4}{*}{\rotatebox[origin=c]{90}{qrels}} & BM25                   & 0.903                  & \textbf{1.000}     \\
 & SPLADE        & \textbf{1.000}        & \textbf{1.000}     \\
 & E5                   & 0.993           & \textbf{1.000}     \\
 & Qwen3                  & 0.960                  &  0.611     \\
\midrule
\multirow{4}{*}{\rotatebox[origin=c]{90}{crels}} & BM25               & 0.833                   & 0.874 \\
 & SPLADE       & \textbf{0.948}         & \textbf{0.922} \\
 & E5                  & 0.931          & 0.862 \\
 & Qwen3                 & 0.933                  &  0.601     \\
\midrule
\multirow{4}{*}{\rotatebox[origin=c]{90}{label\_rels}} & BM25          & 0.083                   & 0.184     \\ 
 & SPLADE          & 0.143                  & 0.204  \\ 
 & E5      & 0.182          & \textbf{0.253} \\
 & Qwen3 & \textbf{0.217}        & 0.206      \\ 
\bottomrule
\end{tabular}
\caption{Performance of the tested retrieval systems for first-stage retrieval. We report performance on the retrieval of chunks according to query relevance (\_qrels), credibility relevance (\_crels), and label relevance (\_label\_rels). \textbf{Boldface} values represent the highest performing system in each category.}\label{tab:retr_res_2}
\end{table}



\begin{table}[ht!]
\small
\centering
\begin{tabular}{@{}lccc@{}}
\toprule
                    & &  \multicolumn{1}{c}{\rabpriv}                      & \multicolumn{1}{c}{\rabpub}       \\
\cmidrule(lr){3-3} \cmidrule(l){4-4}
                         & &  MRR@10                & MRR@10   \\ 
\midrule
\multirow{3}{*}{\rotatebox[origin=c]{90}{BM25}} & qrels        & \textbf{0.913}   & 0.972 \\
 & crels          & \textbf{0.865}       & \textbf{1.000}     \\
 & label\_rels        & \textbf{0.107}       & \textbf{0.248}  \\ \midrule
\multirow{3}{*}{\rotatebox[origin=c]{90}{Qwen3}} & qrels        & \textbf{1.000}      & \textbf{0.942} \\
 & crels          & \textbf{0.960}       & \textbf{0.913} \\
 & label\_rels       & \textbf{0.330}       & \textbf{0.323} \\ \bottomrule
\end{tabular}
\caption{Performance of the BM25 and Qwen3 models with second-stage reranking using the MonoT5 model. \textbf{Boldface} values represent improved performance compared to the corresponding first-stage performance.}\label{tab:rerank_2}
\end{table}

\end{document}